\documentclass[prl,twocolumn,showpacs,amsmath,amssymb,superscriptaddress]{revtex4-1}
\usepackage{dcolumn}
\usepackage{bm,graphicx}
\usepackage{etoolbox}
\usepackage{url}
\usepackage[colorlinks]{hyperref}
\usepackage{breakurl}
\usepackage{color}

\begin{document}

\title{Magneto-optical signature of massless Kane electrons in Cd$_3$As$_2$}

\author{A.~Akrap}
\affiliation{DQMP, University of Geneva, CH-1211 Geneva 4, Switzerland}

\author{M.~Hakl}
\affiliation{LNCMI, CNRS-UGA-UPS-INSA, 25, avenue des Martyrs, 38042 Grenoble, France}

\author{S.~Tchoumakov}
\affiliation{LPS, Univ. Paris-Sud, Univ. Paris-Saclay, CNRS UMR 8502, 91405 Orsay, France}

\author{I.~Crassee}
\affiliation{GAP-Biophotonics, University of Geneva, CH-1211 Geneva 4, Switzerland}

\author{J.~Kuba}

\affiliation{LNCMI, CNRS-UGA-UPS-INSA, 25, avenue des Martyrs, 38042 Grenoble, France}
\affiliation{CEITEC BUT, Brno University of Technology, 616 00 Brno, Czech Republic}

\author{M.~O.~Goerbig}

\affiliation{LPS, Univ. Paris-Sud, Univ. Paris-Saclay, CNRS UMR 8502, 91405 Orsay, France}

\author{C.~C.~Homes}
\affiliation{CMPMS, Brookhaven National Laboratory, Upton, New York 11973, USA}

\author{O.~Caha}
\affiliation{CEITEC MU, Masaryk University, Faculty of Science, 61137 Brno, Czech Republic}

\author{J.~Nov\'ak}
\affiliation{CEITEC MU, Masaryk University, Faculty of Science, 61137 Brno, Czech Republic}

\author{F.~Teppe}
\affiliation{Laboratoire Charles Coulomb, CNRS, Universit\'{e} Montpellier, 34095 Montpellier, France}

\author{W.~Desrat}
\affiliation{Laboratoire Charles Coulomb, CNRS, Universit\'{e} Montpellier, 34095 Montpellier, France}

\author{S.~Koohpayeh}
\affiliation{The Institute for Quantum Matter, The Johns Hopkins University, Baltimore, Maryland 21218, USA}

\author{L.~Wu}
\affiliation{The Institute for Quantum Matter, The Johns Hopkins University, Baltimore, Maryland 21218, USA}
\affiliation{Department of Physics, University of California, Berkeley, CA 94720 USA}

\author{N.~P.~Armitage}
\affiliation{The Institute for Quantum Matter, The Johns Hopkins University, Baltimore, Maryland 21218, USA}

\author{A.~Nateprov}
\affiliation{Institute of Applied Physics, Academy of Sciences of Moldova, MD-2028 Chisinau, Moldova}

\author{E.~Arushanov}
\affiliation{Institute of Applied Physics, Academy of Sciences of Moldova, MD-2028 Chisinau, Moldova}

\author{Q.~D.~Gibson}
\affiliation{Department of Chemistry, Princeton University, Princeton, New Jersey 08544, USA}

\author{R.~J.~Cava}
\affiliation{Department of Chemistry, Princeton University, Princeton, New Jersey 08544, USA}

\author{D.~van~der~Marel}
\affiliation{DQMP, University of Geneva, CH-1211 Geneva 4, Switzerland}

\author{B.~A.~Piot}
\affiliation{LNCMI, CNRS-UGA-UPS-INSA, 25, avenue des Martyrs, 38042 Grenoble, France}

\author{C.~Faugeras}
\affiliation{LNCMI, CNRS-UGA-UPS-INSA, 25, avenue des Martyrs, 38042 Grenoble, France}

\author{G.~Martinez}
\affiliation{LNCMI, CNRS-UGA-UPS-INSA, 25, avenue des Martyrs, 38042 Grenoble, France}

\author{M.~Potemski}
\affiliation{LNCMI, CNRS-UGA-UPS-INSA, 25, avenue des Martyrs, 38042 Grenoble, France}

\author{M.~Orlita}\email{milan.orlita@lncmi.cnrs.fr}
\affiliation{LNCMI, CNRS-UGA-UPS-INSA, 25, avenue des Martyrs, 38042 Grenoble, France}
\affiliation{Institute of Physics, Charles University in Prague, 12116 Prague, Czech Republic}


\begin{abstract}
We report on optical reflectivity experiments performed on Cd$_3$As$_2$ over a broad range of photon energies and magnetic
fields. The observed response clearly indicates the presence of 3D massless charge carriers. The specific
cyclotron resonance absorption in the quantum limit implies that we are probing massless Kane electrons
rather than symmetry-protected 3D Dirac particles. The latter may appear at a smaller energy scale and
are not directly observed in our infrared experiments.
\end{abstract}


\maketitle

Cadmium arsenide (Cd$_3$As$_2$) has recently been identified~\cite{LiuNatureMater14,NeupaneNatureComm14,*BorisenkoPRL14,JeonNatureMater14} as the premier 3D topological Dirac semimetal stable under ambient conditions, thus inspiring a renewed interest in the electronic properties of this widely investigated compound~\cite{RosenmanJPCS69,Bodnar77,GeltenSSC80,*AubinPRB81,*ArushanovPCGC80,*ArushanovPCGC92,SchleijpeIJIMW84}. The current consensus is that the electronic bands of Cd$_3$As$_2$ comprise a single pair of symmetry-protected 3D Dirac nodes located in the vicinity of the $\Gamma$ point of the Brillouin zone. Nevertheless, the exact location, size, anisotropy and tilt of these conical bands still remain a puzzle.
Most strikingly, ARPES studies imply cones extending over a few hundred meV~\cite{NeupaneNatureComm14,BorisenkoPRL14} or even eV~\cite{LiuNatureMater14}. In contrast, the band inversion estimated in STM/STS experiments~\cite{JeonNatureMater14}, in line with recent and past theoretical modeling~\cite{Bodnar77,WangPRB13}, invokes Dirac cones which extend over an order-of-magnitude smaller energy range.

In this Letter, we clarify the controversies on the electronic bands of Cd$_3$As$_2$. Based on our magneto-optical experiments, we argue that the
band structure may in fact include two types of conical features, one spread over the large, and the second on the small energy scale. The widely extended conical band results from the standard Kane model~\cite{KaneJPCS57,Kacmanppsb71} applied to a semiconductor with a nearly vanishing band gap, it is not symmetry-protected and it hosts carriers that are referred to as massless Kane electrons~\cite{OrlitaNaturePhys14}. The symmetry-protected Dirac cones, if present in Cd$_3$As$_2$ at all, may only appear  on a much smaller energy scale, in contrast to conclusions of ARPES studies~\cite{LiuNatureMater14,NeupaneNatureComm14,*BorisenkoPRL14}, but in line with the STM/STS data~\cite{JeonNatureMater14}.

\begin{figure}[t]
\includegraphics[trim = 0mm 0mm 0mm 0mm, clip=true, width=8cm]{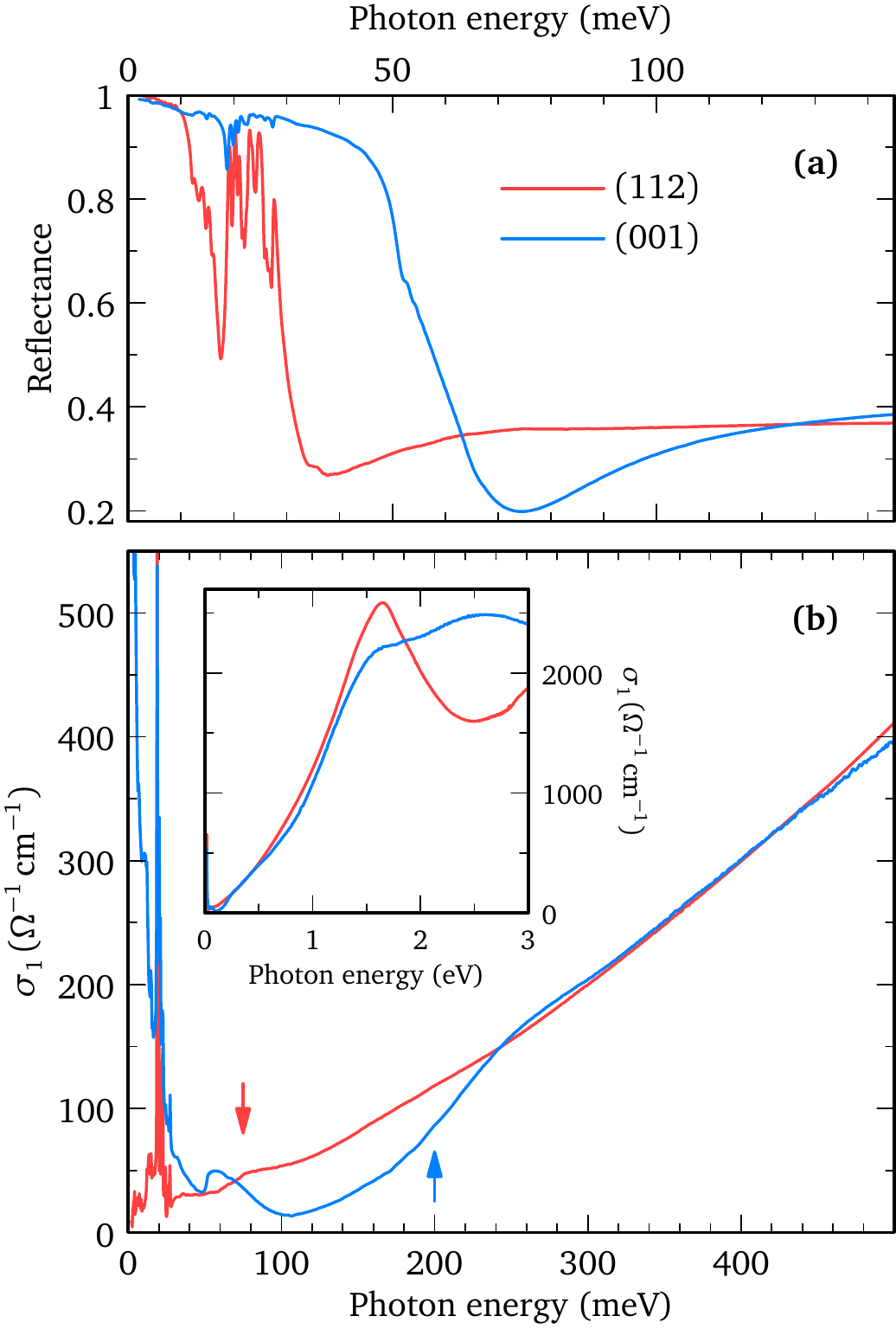}
\caption{\label{Conductivity001&112} (color online) (a) The reflectance of both studied samples at 5~K and $B=0$ dominated by sharp plasma edges and phonon response.
For a purely conical band structure, the approximate ratio of plasma frequencies, $\omega^{001}_p/\omega^{112}_p\approx 2$, implies $2\times$ larger $E_F$, and therefore, $8\times$ higher carrier density ($n\propto E_F^3$) for the (001)-oriented sample. (b) The optical conductivity with a nearly linear dependence on the photon energy. Inset: optical conductivity in a wide energy range.}
\end{figure}

Our experiments were performed on two bulk Cd$_3$As$_2$ $n$-doped samples with facets oriented using $x$-ray scattering
experiments~\cite{SM}. The sample with the (001) facet ($\sim$1$\times$2~mm$^2$) was cut and polished from a bulk crystal prepared using the
technique described in Ref.~\onlinecite{AliIC14}. The (112)-oriented sample ($\sim$2$\times$4~mm$^2$) was grown using a thermal evaporation method~\cite{WeberAPL15}.

The reflectance experiments were performed using standard Fourier transform spectroscopy.
In experiments at $B=0$, the reflectivity was measured in the range of 0.003-3~eV using an in situ overcoating technique.
Ellipsometry was employed to determine the dielectric function between 0.5 and 4~eV.
To measure reflectivity in magnetic fields, a resistive coil was used, with a liquid helium bath cryostat at $T=1.8$~K,
and samples placed in a low-pressure helium exchange gas.

The reflectance spectra taken at $B=0$ and $T=5$~K are plotted in Fig.~\ref{Conductivity001&112}a. The observation of characteristic plasma edges
indicates the presence of free charge carriers. Sharp phonon response is clearly observed around 15-30~meV.
For both samples, the optical conductivity (Fig.~\ref{Conductivity001&112}b) was extracted using the Kramers-Kronig analysis of the reflectivity data, with the phase anchored by the high-frequency ellipsometry data. At first glance, the optical conductivity increases linearly with the photon frequency, which is consistent with the response
of 3D massless particles~\cite{TimuskPRB13,OrlitaNaturePhys14}. A closer inspection reveals a slight superlinear increase, which is discussed
later on (cf. similar recent data in Ref.~\onlinecite{NeubauerPRB16}).

At low energies, the optical conductivity reflects different electron densities in the studied samples, which implies
different onsets of interband absorption due to Pauli blocking. These onsets were identified as inflection points in the conductivity
spectra and marked by vertical arrows in Fig.~\ref{Conductivity001&112}b. In an ideal conical band, such an onset corresponds to twice the Fermi
energy. In Cd$_3$As$_2$, these onsets are closer to $E_F$, due to the electron-hole asymmetry, as confirmed
\emph{a posteriori} by our data analysis. For 3D massless particles, the plasma frequency is predicted to scale linearly with the Fermi
energy, $\omega_p\propto E_F$~\cite{DasSarmaPRL09}.
Indeed, for our two samples, the onset energy ratio approaches 2 and matches the plasma frequency ratio. At higher photon energies, the optical conductivity
indicates a fairly high isotropy of Cd$_3$As$_2$, with the first signs of anisotropy appearing
above 0.5~eV (inset of Fig.~\ref{Conductivity001&112}b).

\begin{figure*}[t]
      \includegraphics[trim = 0mm 0mm 0mm 0mm, clip=true, width=16cm]{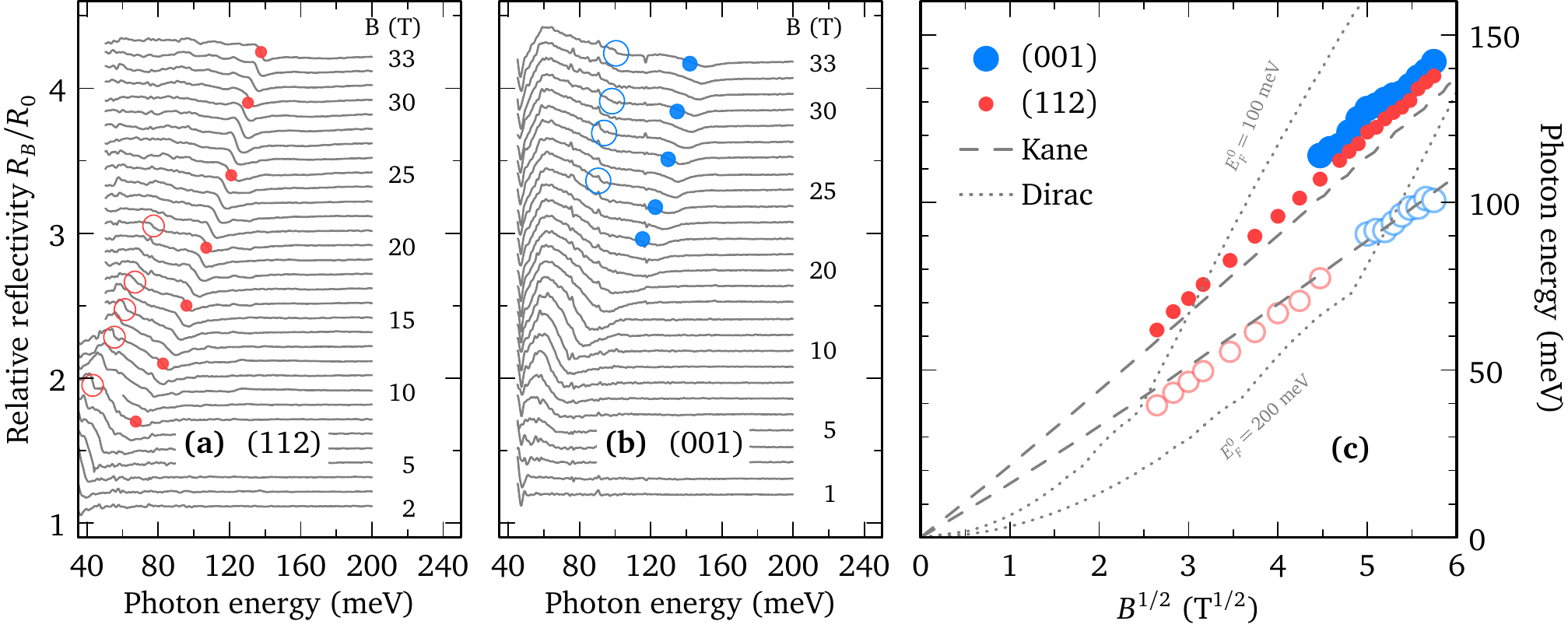}
      \caption{\label{Inter-LL-transitions} (color online)  Relative magneto-reflectivity spectra of the (112)- and (001)-oriented sample measured at $T=1.8$~K
      in (a) and (b), respectively. The observed resonances are marked by full (open) circles for dominant (weak, likely present) CR modes.
      Their $\sqrt{B}$ dependence is shown in the part (c), together with the theoretical curves based on the Dirac (dotted) and gapless Kane (dashed) models.
      The former curves were calculated for the velocity $10^6$~m/s and the zero-field Fermi energies $E^0_F=100$ and 200 meV; the kinks appear due to field-dependence of $E_F$~\cite{SM}.}
\end{figure*}

The presence of massless electrons in Cd$_3$As$_2$ is also visible in our high-field magneto-reflectivity data (Fig.~\ref{Inter-LL-transitions}).
At low $B$, the plasma edge undergoes a clear splitting in both samples~\cite{SM}, which is fully consistent with previously published
experiments~\cite{SchleijpeIJIMW84} and which can be described using a classical magneto-plasma theory~\cite{PalikRPP70} with the quasi-classical
cyclotron frequency $\omega_c$ linear in $B$. At high $B$, for $\omega_c \gg \omega_p$, a well-defined cyclotron resonance (CR) mode appears,
likely accompanied by a weaker satellite line (full and open circles in Figs.~\ref{Inter-LL-transitions}a,b, respectively).
These modes, determined as the corresponding inflection points in the spectra~\cite{SM}, follow a $\sqrt{B}$ dependence (Fig.~\ref{Inter-LL-transitions}c)
-- a hallmark of the characteristic spacing of Landau levels (LLs) of massless particles.

The high-field CR absorption confirms the isotropic nature of the conical band structure consistent with the zero-field response.
A closer look at the data obtained on samples with differently oriented facets reveals a small difference in the measured CR energies,
$\omega_c^{001}/\omega_c^{112}\sim1.03$ (Fig.~\ref{Inter-LL-transitions}c). This difference, comparable with the experimental accuracy of the
CR energy, indicates a weak anisotropy, $v_\perp/v_\|\approx 0.9$, expressed using the in-plane $v_\|$ and out-of-plane $v_\perp$ velocity~\cite{SM}.

Let us compare, first on a \emph{qualitative} level, our experimental findings with the electronic band structure expected within the semi-empirical model proposed
for Cd$_3$As$_2$ by Bodnar~\cite{Bodnar77}. This model extends the standard Kane model~\cite{KaneJPCS57},
widely applied to describe the band structure of zinc-blende (cubic) semiconductors at the $\Gamma$ point and based on the exact
diagonalization of the $\mathbf{k}\cdot \mathbf{p}$ Hamiltonian, using the finite basis of $s$ and $p$ valence states.
Bodnar retained the standard Kane parameters (band gap $E_g$, interband matrix element $P$ and spin-orbit coupling $\Delta$),
while additionally introducing the crystal field splitting $\delta$~\cite{KildalPRB74}.
The latter parameter $\delta$ reflects the tetragonal symmetry of Cd$_3$As$_2$ and splits the 4-dimensional $\Gamma_8$
representation, which in cubic semiconductors corresponds to heavy and light hole bands, degenerate at the $\Gamma$ point.

\begin{figure*}
\includegraphics[trim = 0mm 0mm 00mm 0mm, clip=true, width=13.5cm]{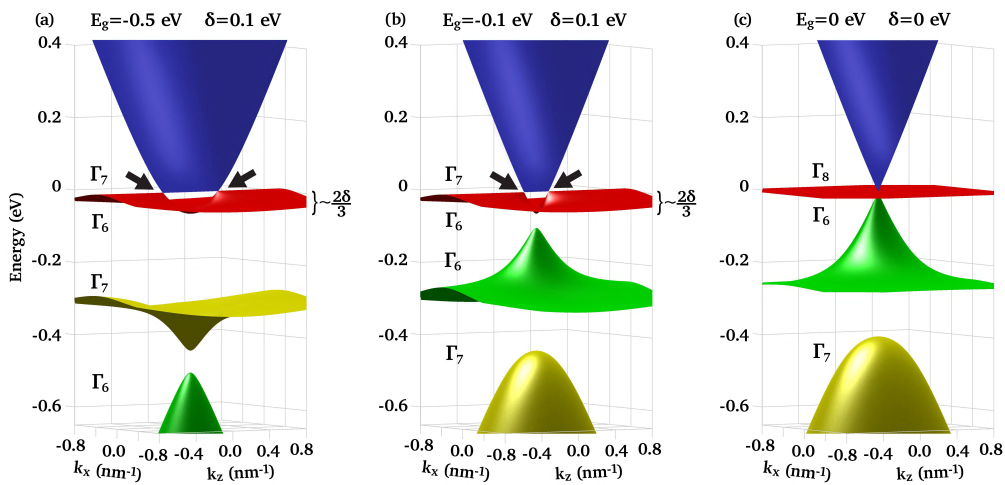}
\caption{\label{Bodnar} The band structure calculated using the Bodnar model (at $k_y\equiv0$) for various $E_g$ and $\delta$ chosen to approach (a) ab initio calculations~\cite{WangPRB13}, (b) the original Bodnar's work~\cite{Bodnar77} and (c) Kane electrons~\cite{OrlitaNaturePhys14}, and for $\Delta=400$~meV and $v=0.93\times10^6$~m/s. Two highly tilted Dirac cones (marked by black arrows) with the size comparable to $\delta$ appear in (a) and (b) at the crossing points
of the upper $\Gamma_7$ and $\Gamma_6$ bands. The widely extended cone of Kane electrons is visible in (c) and partly also (b).}
\end{figure*}

The model allows for the existence of two different branches of 3D massless electrons, both having their
charge neutrality point at the Fermi energy in an undoped system. For $E_g<0$ (\emph{i.e.}, in the inverted band gap regime) and $\delta>0$, the $\Gamma_6$ and $\Gamma_7$
bands, which originate in the split $\Gamma_8$ representation, form  an avoided crossing in all momentum directions, except
the tetragonal ($z$) axis, where such an anticrossing is forbidden by symmetry (due to the $C_4$ rotation axis~\cite{YangNatureComm14}).
As a consequence, two strongly tilted and anisotropic 3D cones are created at $k_z^D=\pm\sqrt{|E_g|\delta/P^2}$
(Fig.~\ref{Bodnar}a,b). This specific behavior was postulated in an earlier work~\cite{Bodnar77}
and the cones were later classified as symmetry-protected Dirac cones~\cite{WangPRB13}.

Another sort of massless particle appears when both the band gap and crystal field splitting vanish.
In this limit, the Bodnar model becomes equivalent to that of massless Kane electrons~\cite{Kacmanppsb71,OrlitaNaturePhys14}
with a single isotropic spin-degenerate cone at the $\Gamma$ point (Fig.~\ref{Bodnar}c). This cone is characterized by a
velocity $v=\sqrt{2/3}P/\hbar$ and, additionally, by the appearance of a flat heavy-hole band. The anisotropy
of this cone may be introduced in the model through an anisotropic matrix element $P$.

Notably, the Bodnar model generically allows for a crossover from massless Dirac to Kane electrons (Fig.~\ref{Bodnar}b). For positive $\delta$ and negative $E_g$ small as compared to $\Delta$, two symmetry-protected Dirac cones are formed at low energies, with the size comparable to $\delta$. These Dirac cones then merge, above the Lifshitz point, into a single conical band. At higher energies, this single cone is well described by the gapless Kane model and thus hosts particles with a linear dispersion -- the so-called massless Kane electrons~\cite{OrlitaNaturePhys14}.

The low anisotropy of the observed optical response is the first indication that one is dealing with a single
cone of Kane electrons (around the $\Gamma$ point) rather than two Dirac cones (located along the tetragonal axis at $\pm k_z^D$).
More direct evidence comes from the qualitative analysis of our high-field CR data.
In the quantum regime, materials with massless electrons imply multimode CR absorption. The individual modes correspond to excitations between adjacent
non-equidistantly spaced LLs of the same spin, analogous to the CR response of graphene~\cite{NeugebauerPRL09}.
The intensity of these modes is related to the number of charge carriers in particular LLs,
which follows their degeneracy and spacing. We have found two such modes in the response of each Cd$_3$As$_2$ sample.

The magneto-reflectivity response of the (112)-oriented sample is dominated by a CR mode, which emerges below $B=10$~T and
remains present up to the highest magnetic field applied, $B=33$~T (Fig.~\ref{Inter-LL-transitions}a). In any system of
either Dirac or Kane massless electrons, this is only possible when the quantum limit is approached and
electrons are excited from the lowest LL in the conduction band. This is because this fundamental level cannot be, in contrast to
higher LLs, depopulated with the increasing $B$.

\begin{figure}[t]
\includegraphics[trim = 0mm 0mm 0mm 0mm, clip=true, width=\linewidth]{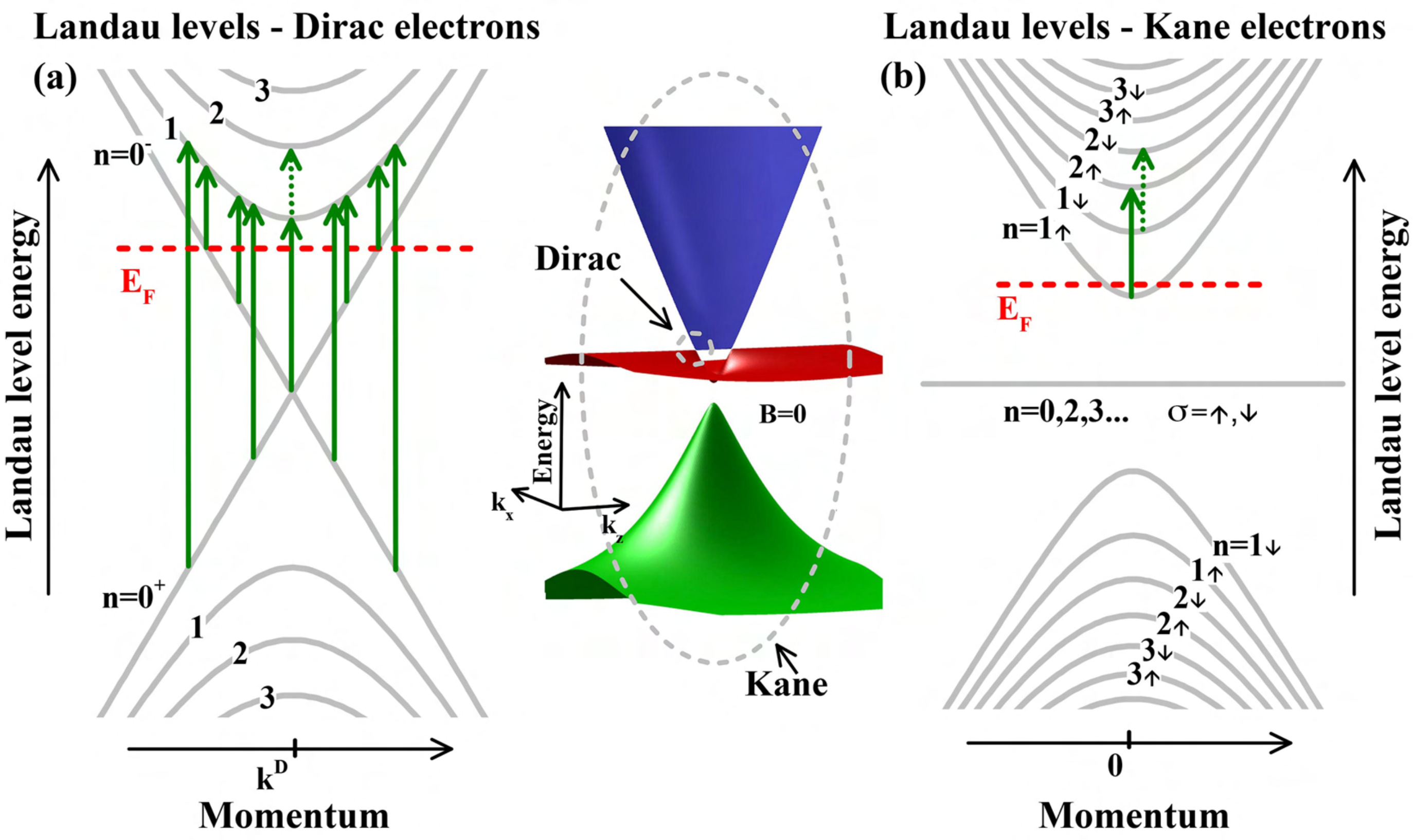}
\caption{\label{LLs} Idealized Landau levels of (a) Dirac and (b) Kane electrons as occupied in the quantum limit and their
relative energy scale in the Bodnar model (central inset).
The $n=0$ LLs typical of Dirac electrons, dispersing linearly with the momentum, are absent for Kane electrons.
The solid arrows show CR modes active in the quantum limit, the dashed arrows those which vanish when this limit is reached.
For details about the crossover from Dirac- to Kane-like LLs see Ref.~\onlinecite{SM}.}
\end{figure}

To test whether 3D Dirac electrons may be at the origin of the observed (isotropic) CR response, let us consider their
LL spectrum, $E^{\mathrm{Dirac}}_n=\pm v\sqrt{2e\hbar B n +\hbar^2 k^2}$. It comprises specific $n=0$ LLs dispersing
linearly with the momentum $k$ along the magnetic field. Each CR mode between adjacent LLs
with $n\geq1$ is characterized by a singularity in the joint density of states due to the band extrema at $k=0$, and therefore, by a strict
$\sqrt{B}$ dependence. For the fundamental CR mode, which is the only active mode in the quantum limit and corresponds
to excitations from the $n=0$ levels, such a singularity is missing because of the linear dispersion in $k$
(Fig.~\ref{LLs}a). Its only characteristic spectral feature is the low-energy absorption edge
at L$_{0\rightarrow1}=\sqrt{v^2 2e\hbar B+E_F^2}-E_F$~\cite{AshbyPRB13}.

Importantly, this low-energy edge does not follow the $\sqrt{B}$ dependence, and moreover, its position
depends on $E_F$. This is in clear disagreement with our experimental findings, where the fundamental CR lines
(full circles in Fig.~\ref{Inter-LL-transitions}c) show a nearly perfect $\sqrt{B}$ dependence and have almost
the same position for both studied samples (despite significantly different $E_F$). The picture of 3D Dirac electrons
thus does not match our data. We illustrate this in Fig.~\ref{Inter-LL-transitions}c by
plotting the L$_{0\rightarrow1}$ energy for two selected zero-field Fermi levels $E_F^0=100$ and 200~meV. These
correspond to the chemical potential estimated from the zero-field optical conductivity in Fig.~\ref{Conductivity001&112}b
for the two investigated specimens. The field dependence of $E_F$  was calculated supposing
a 3D Dirac LL spectrum (see Ref.~\cite{SM} for further details).
The plotted L$_{0\rightarrow1}$ curves clearly deviate from the $\sqrt{B}$ dependence and thus cannot be used for fitting
by any variation of the velocity parameter (fixed at $10^6$~m/s in the plot).

In contrast, $n=0$ Landau levels dispersing linearly in $k$
are absent in systems with Kane electrons,
$E^{\mathrm{Kane}}_{\zeta,n,\sigma}=\zeta v\sqrt{e\hbar B (2n-1+\sigma)+\hbar^2 k^2}$, where $\zeta=-1,0,1$ and $\sigma=\pm1/2$~\cite{OrlitaNaturePhys14}.
As a result, all CR modes of Landau-quantized massless Kane electrons, including the fundamental one, follow a $\sqrt{B}$ dependence.
Notably, in an electron-doped system, the Fermi energy never drops below the $n=1^\uparrow$ level
and the fundamental mode thus corresponds to $n=1^\uparrow \rightarrow 2^\uparrow$ (Fig.~\ref{LLs}b).

After these qualitative arguments, we now proceed with the \emph{quantitative} analysis of our data, using the simplest gapless Kane model~\cite{OrlitaNaturePhys14}.
This model implies only two material parameters, $v$ and $\Delta$. The velocity parameter was taken after Bodnar, $v=0.93\times10^6$~m/s, in perfect
agreement with a value recently deduced from STM/STS measurements~\cite{JeonNatureMater14}, $0.94\times10^6$~m/s, determined for a single conical band at the $\Gamma$ point. The strength of the spin-orbit coupling was set to $\Delta=400$~meV, the value known for InAs. This choice is justified by the simple fact that the strength of the spin-orbit interaction
in most semiconductors is predominantly governed by the anion atom, which in this case is arsenic. Lower
values might also be considered (e.g., $\Delta=270$~meV~\cite{Bodnar77}), at the expense of a slightly increased electron-hole asymmetry in the model.

Figure~\ref{Inter-LL-transitions}c shows the theoretical positions of the two CR modes expected to be active in the vicinity of the quantum limit of
massless Kane electrons (dashed lines). These modes correspond to excitations between the two lowest spin-split Landau levels in the conduction band (from $n=1$ to 2),
as denoted by vertical arrows in Fig.~\ref{LLs}b.
The Kane model thus fits well the experimental points for the dominant CR line
as well as for its weaker satellite, using only the above estimates for $v$ and $\Delta$, introducing
no additional fitting parameters. Consistent with this picture, the satellite line is gradually suppressed with $B$
in the lower-doped (112)-oriented sample, when the system enters the quantum limit.

Although the Kane model explains our data fairly well, due to the tetragonal nature of Cd$_3$As$_2$ this model cannot be valid down to arbitrarily low energies.
The comparison with the full Bodnar model~\cite{SM} gives rough estimates (upper limits) for
the band gap and crystal field splitting: $E_g=-30$~meV and $\delta=30$~meV. This choice of parameters is consistent with a band structure similar to the one in Fig.~\ref{Bodnar}b, where 3D Dirac electrons are present but are limited to low energies. This confirms that the  (magneto-)optical response studied in our experiments is dominated by the massless Kane and not Dirac electrons.

Having concluded the presence of Kane electrons in Cd$_3$As$_2$, we return to the zero-field data (Fig.~\ref{Conductivity001&112}b).
The optical conductivity in an ideal undoped system of Kane electrons (for $\Delta$$\,\rightarrow\,$$\infty$) may be written
as $\mathrm{Re}\{\sigma^{\mathrm{Kane}} (\omega) \}= (13/12) \omega e^2/(4\pi\hbar v)$~\cite{OrlitaNaturePhys14}.
The dominant contribution to this interband absorption stems from excitations from the flat band (red in Fig.~\ref{Bodnar}c)
to the upper cone (blue). This may be viewed as interband absorption in the cone with an extreme electron-hole asymmetry and confirms our assignment of the
onsets in optical conductivity to $E_F$  (and not $2E_F$). The contribution from the lower (green) cone is weaker mainly due to the $8\times$ smaller joint density of states.

In our case, however, the lower cone flattens significantly when approaching the energy of $-\Delta$ (Figs.~\ref{Bodnar}b,c) and the conductivity should be,
in the first approximation, corrected by an additional term~\cite{SM}, which nearly doubles its slope for $\hbar\omega>\Delta$:
$$
\mathrm{Re}\{\sigma^{\mathrm{Cd_3As_2}}(\omega)\}\approx  \frac{\omega e^2}{4\pi\hbar v}\left[\frac{13}{12}+\theta(\hbar\omega-\Delta)
\left(1-\frac{\Delta}{\hbar\omega}\right)^2\right],
$$
where $\theta(x)$ is the Heaviside function. This may explain the superlinear increase (upturn) in the spectra around 0.5~eV (Fig.~\ref{Conductivity001&112}b) as a simple band-effect,
providing thus an alternative to the recent interpretation in terms of self-energy corrections within the Dirac picture~\cite{NeubauerPRB16}.
One may expect such corrections, as well as enhanced interaction effects,
rather at high energies ($1-2$ eV) where the maximum in the optical conductivity
hints at an enlarged density of states.

On the other hand, the above expression for conductivity still underestimates the total observed absorption strength. Only semi-quantitative
agreement is thus achieved, which points towards other contributions to conductivity. These include, for instance, enhanced absorption
at low energies ($\hbar\omega<\Delta$) due to gradually flattened lower cone, as well as excitations into the second conduction band,
not included in the Bodnar/Kane model, but expected at energies around $\sim$0.5~eV above the flat band~\cite{WangPRB13,JeonNatureMater14}.

In summary, we have studied bulk Cd$_3$As$_2$ by means of optical magneto-spectroscopy, unambiguously showing the presence of 3D massless particles.
We conclude that within the investigated range of frequencies and magnetic fields, the observed response is due to massless Kane electrons,
the presence of which is fully consistent with the Bodnar model elaborated for this material in the past. This model is, in our opinion,
also applicable to study the symmetry-protected 3D Dirac electrons. These may appear in Cd$_3$As$_2$ at a small energy scale
given by the crystal field splitting (a few ten meV at most). Our finding contradicts the conclusions of ARPES
studies~\cite{LiuNatureMater14,NeupaneNatureComm14,*BorisenkoPRL14}, in which the observed conical feature
extending over a few hundred meV was interpreted in terms of symmetry-protected Dirac particles. Nevertheless,
the identified massless Kane electrons should exhibit some properties that are typical of truly relativistic particles, with
the Klein tunnelling as a prominent example~\cite{KatsnelsonNaturePhys06,LiangNatureMater15}.

\begin{acknowledgments}
This work was supported by ERC MOMB (No. 320590), TWINFUSYON (No. 692034), Lia TeraMIR, TERASENS and by MEYS CEITEC 2020 (No. LQ1601) projects.
Support was also received from the U.S. Department of Energy, Office of Basic Energy Sciences, MSE Division und
Contract No. DE-SC0012704 (C.C.H.) and DE-FG02-08ER46544 (S.K., L.W. and N.P.A.). A.A. acknowledges funding from the Ambizione grant of the Swiss NSF.
Work at Princeton University was supported by the ARO MURI on topological insulators (W911NF-12-0461). The authors acknowledge discussions with D.~M. Basko, M.~Civelli, and W.-L. Lee.
\end{acknowledgments}

%

\newpage
\pagenumbering{gobble}

\begin{figure}[htp]
\includegraphics[page=1,trim = 17mm 17mm 17mm 17mm, width=1.0\textwidth,height=1.0\textheight]{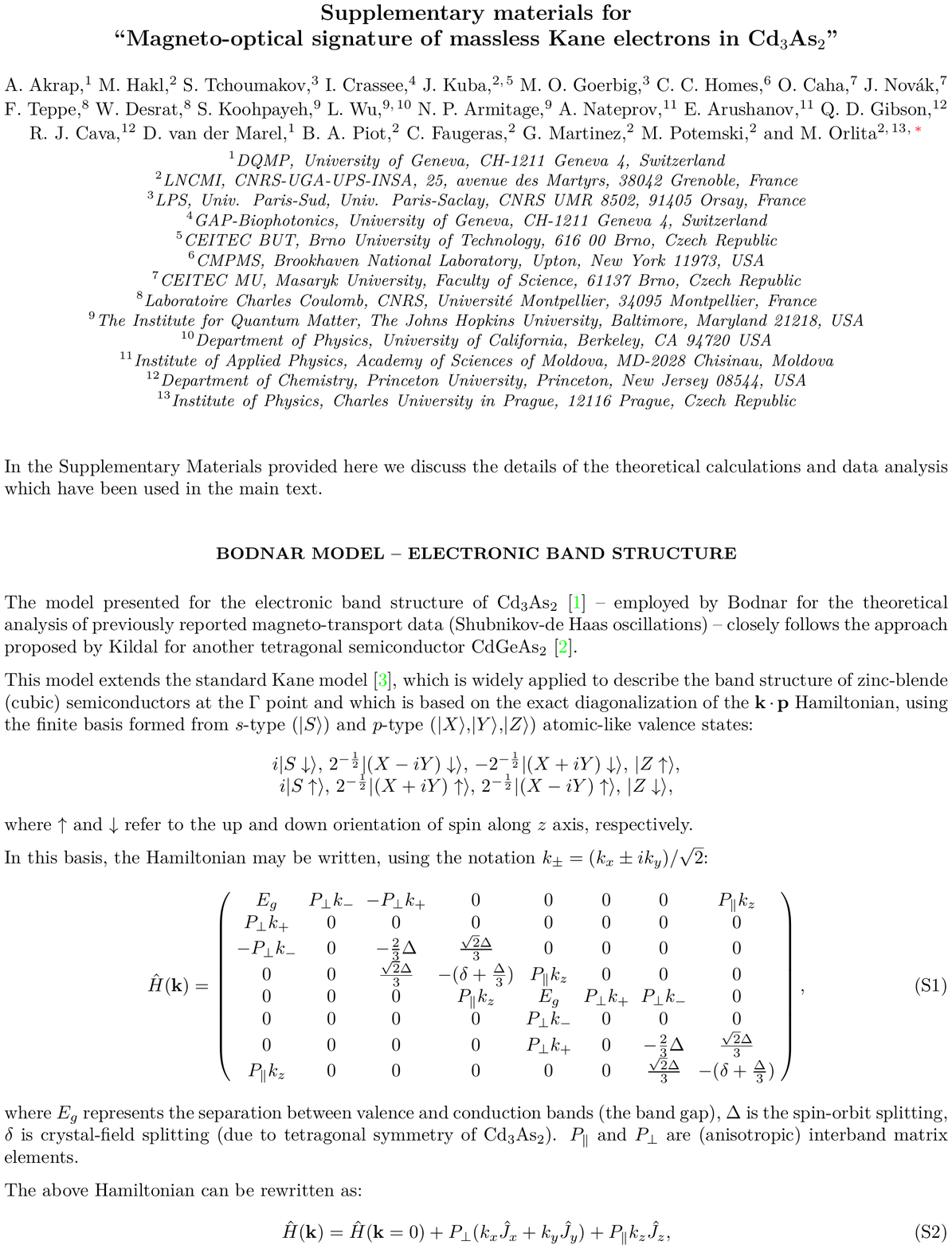}

\end{figure}

\newpage

\begin{figure}[htp]
  \includegraphics[page=2,trim = 17mm 17mm 17mm 17mm, width=1.0\textwidth,height=1.0\textheight]{SI_resubmission.pdf}

\end{figure}

\newpage

\begin{figure}[htp]
  \includegraphics[page=3,trim = 17mm 17mm 17mm 17mm, width=1.0\textwidth,height=1.0\textheight]{SI_resubmission.pdf}

\end{figure}

\newpage

\begin{figure}[htp]
  \includegraphics[page=4,trim = 17mm 17mm 17mm 17mm, width=1.0\textwidth,height=1.0\textheight]{SI_resubmission.pdf}

\end{figure}

\newpage

\begin{figure}[htp]
  \includegraphics[page=5,trim = 17mm 17mm 17mm 17mm, width=1.0\textwidth,height=1.0\textheight]{SI_resubmission.pdf}

\end{figure}

\newpage

\begin{figure}[htp]
  \includegraphics[page=6,trim = 17mm 17mm 17mm 17mm, width=1.0\textwidth,height=1.0\textheight]{SI_resubmission.pdf}

\end{figure}

\newpage

\begin{figure}[htp]
  \includegraphics[page=7,trim = 17mm 17mm 17mm 17mm, width=1.0\textwidth,height=1.0\textheight]{SI_resubmission.pdf}

\end{figure}

\newpage

\begin{figure}[htp]
  \includegraphics[page=8,trim = 17mm 17mm 17mm 17mm, width=1.0\textwidth,height=1.0\textheight]{SI_resubmission.pdf}

\end{figure}

\newpage

\begin{figure}[htp]
  \includegraphics[page=9,trim = 17mm 17mm 17mm 17mm, width=1.0\textwidth,height=1.0\textheight]{SI_resubmission.pdf}

\end{figure}

\newpage

\begin{figure}[htp]
  \includegraphics[page=10,trim = 17mm 17mm 17mm 17mm, width=1.0\textwidth,height=1.0\textheight]{SI_resubmission.pdf}

\end{figure}

\newpage

\begin{figure}[htp]
  \includegraphics[page=11,trim = 17mm 17mm 17mm 17mm, width=1.0\textwidth,height=1.0\textheight]{SI_resubmission.pdf}

\end{figure}

\newpage

\begin{figure}[htp]
  \includegraphics[page=12,trim = 17mm 17mm 17mm 17mm, width=1.0\textwidth,height=1.0\textheight]{SI_resubmission.pdf}

\end{figure}

\newpage

\begin{figure}[htp]
  \includegraphics[page=13,trim = 17mm 17mm 17mm 17mm, width=1.0\textwidth,height=1.0\textheight]{SI_resubmission.pdf}

\end{figure}

\newpage

\begin{figure}[htp]
  \includegraphics[page=14,trim = 17mm 17mm 17mm 17mm, width=1.0\textwidth,height=1.0\textheight]{SI_resubmission.pdf}

\end{figure}

\newpage

\begin{figure}[htp]
  \includegraphics[page=15,trim = 17mm 17mm 17mm 17mm, width=1.0\textwidth,height=1.0\textheight]{SI_resubmission.pdf}

\end{figure}

\newpage

\begin{figure}[htp]
 \includegraphics[page=16,trim = 17mm 17mm 17mm 17mm, width=1.0\textwidth,height=1.0\textheight]{SI_resubmission.pdf}

\end{figure}

\newpage

\begin{figure}[htp]
  \includegraphics[page=17,trim = 17mm 17mm 17mm 17mm, width=1.0\textwidth,height=1.0\textheight]{SI_resubmission.pdf}
  
\end{figure}

\end{document}